\documentclass{article}
\usepackage{spconf,amsmath,graphicx}
\usepackage{algorithmic}
\usepackage{algorithm}
\usepackage{booktabs}
\usepackage{url}

\title{ACE-VC: Adaptive and Controllable Voice Conversion using Explicitly Disentangled Self-supervised Speech Representations}

\name{$^*$Shehzeen Hussain$^1$ \qquad $^*$Paarth Neekhara$^1$\thanks{* Equal contribution. Work performed as interns at NVIDIA} \qquad Jocelyn Huang$^2$ \qquad Jason Li$^2$ \qquad Boris Ginsburg$^2$ }
\address{$^{1}$University of California, San Diego, CA, USA \\
      {$^{2}$}NVIDIA Corporation, Santa Clara, CA, USA }
\begin{document}

\maketitle
\begin{abstract}
In this work, we propose a zero-shot voice conversion method using speech representations trained with self-supervised learning.
First, we develop a multi-task model to decompose a speech utterance into features such as linguistic content, speaker characteristics, and speaking style. 
To disentangle content and speaker representations, we propose a training strategy based on Siamese networks that encourages similarity between the content representations of the original and pitch-shifted audio.
Next, we develop a synthesis model with pitch and duration predictors that can effectively reconstruct the speech signal from its decomposed representation. 
Our framework allows controllable and speaker-adaptive synthesis to perform zero-shot any-to-any voice conversion achieving state-of-the-art results on metrics evaluating speaker similarity, intelligibility, and naturalness.
Using just $10$ seconds of data for a target speaker, our framework can perform voice swapping and achieves a speaker verification EER of $5.5\%$ for seen speakers and $8.4\%$ for unseen speakers.~\footnote{Audio and Code: \url{https://paarthneekhara.github.io/ace}}


\end{abstract}
\section{Introduction}
\label{sec:intro}

Voice conversion is the task of modifying an utterance from a source speaker to match the vocal qualities of the target speaker. 
While traditional voice conversion systems~\cite{saito2011one,mohammadi2017overview} rely on parallel training data with multiple speakers saying the same sentence,
there has been a recent surge in voice conversion systems trained on non-parallel multi-speaker datasets~\cite{chou2019one,qian2019autovc,choi2021neural,casanova2022yourtts}.
The key idea behind non-parallel voice conversion systems is disentangling speech into representations describing the linguistic content and speaker characteristics.
Synthesizing speech from these disentangled features allows voice conversion by swapping the speaker embedding of a given utterance with a target speaker.



To disentangle content and speaker information from a speech signal for voice conversion, some prior work~\cite{sun2016phonetic,tian2018average} have utilized pre-trained automatic speech recognition (ASR) and speaker verification (SV) models.
The predicted text or the phonetic posteriogram (PPG) extracted using the ASR model is considered the content representation, while the embedding derived from the SV model is considered the speaker representation. 
While this approach has shown promise in voice conversion, it has a few limitations --- 1) Such a system is sensitive to ASR errors which can cause mispronunciation of certain words or inaccurate conversion.  2) Text and PPG do not capture all linguistic features such as accent, expressions and speaker-independent style resulting in neutral-sounding synthesized speech.



Recently, researchers~\cite{lakhotia2021generative,polyak2021speech,lin2021fragmentvc,huang2022s3prl} have proposed techniques to synthesize speech from the outputs of models trained using self-supervised learning (SSL). While the SSL representations are highly correlated with phonetic information, they do not necessarily disentangle speaker and linguistic content and can be effective for both speech recognition~\cite{wav2vec2,gulati2020conformer} and speaker verification~\cite{hussain2022multi}.
Therefore, to extract the disentangled linguistic content, SSL representations are often quantized to obtain \textit{pseudo-text} from a speech utterance. 
The main limitation of these methods is that quantizing SSL representations is a lossy compression, which leads to sub-optimal reconstruction quality. 
Moreover, there is no guarantee that speaker information is absent in the quantized representations since disentanglement is not explicitly enforced in such a setup.

To address the above limitations in voice conversion systems, we develop a Speech Representation Extractor (SRE) to more effectively disentangle speaker information and linguistic content from the SSL representations of a given speech utterance. 
Our proposed SRE relies on a Conformer-SSL model~\cite{gulati2020conformer} that is trained in a multi-task manner on two downstream tasks --- automatic speech recognition and speaker verification. 
To disentangle the content and speaker representations during training, we synthetically modify the voice of a given speech using pitch-shift transform and then enforce similarity between the content embeddings of the original and transformed audio using a Siamese setup~\cite{bromley1993signature} with cosine-similarity loss. 
In contrast to past works~\cite{chou2019one,qian2019autovc,choi2021neural,casanova2022yourtts,huang2022s3prl}, our proposed technique uses a single neural network (Conformer-SSL) backbone to derive both content and speaker information. We enforce feature disentanglement without compressing or quantizing the content representation.



Finally, we  develop a synthesis network to reconstruct speech from the SRE representations. Our synthesis network incorporates learnable intermediate models that can predict the fundamental frequency and token duration from both content and speaker representations. 
Our synthesis framework enables two functionalities for generated speech -- either mimicking the prosody and rhythm of the source speech using ground truth pitch and duration or adapting as per the target speaker embedding using predicted pitch and duration. 
Our voice conversion framework achieves state-of-the-art results in zero-shot voice conversion for both seen and unseen speakers on metrics evaluating speaker similarity, intelligibility and naturalness of synthesized speech.

\section{Methodology}
\label{sec:method}
Our framework for voice conversion consists of three major components that are trained separately --- 1) An SSL based Speech Representation Extractor (SRE) 2) an upstream Mel-Spectrogram synthesizer and 3) a HiFi-GAN vocoder~\cite{kong2020hifi}.  


\subsection{Speech Representation Extractor (SRE)}
\label{sec:sre}
The goal of the SRE is to extract disentangled speaker and content representations from a given audio waveform. To this end, we utilize the Conformer model~\cite{gulati2020conformer} trained in a self-supervised manner as the backbone of our framework. 
The Conformer architecture combines convolutions and multi-head self-attention blocks 
resulting in a computationally efficient network that can model both local and long-term dependencies.
The network operates on mel-spectrogram representation of audio and is trained to predict the quantized values of the masked inputs using a contrastive loss. 
Given a mel-spectrogram $x=(x_1,\dots x_T$) as a sequence of frames, the Conformer $E$ outputs a sequence of vectors $z=E(x)= (z_1 \dots z_{T'} )$. 

To extract the linguistic content and speaker representations from our pre-trained Conformer SRE backbone, we add two randomly initialized downstream heads and perform multi-task training for speech recognition and speaker verification: 

\noindent \textbf{1) Speech recognition:} For speech recognition, the downstream head comprises two linear layers. The first linear layer maps the SRE backbone outputs to a sequence of content embedding $z_c = f_{\theta_{c1}} (z) = {z_c}_1 \dots {z_c}_{T'} $. The second linear layer followed by softmax, maps the content embedding at each time-step to probability scores over the language tokens, that is, $p_c = f_{\theta_{c2}} (z_c) = {p_c}_1 \dots {p_c}_{T'} $. The speech recognition downstream head is trained using CTC loss~\cite{graves2006connectionist} on audio-text pairs $(x, y_c)$. That is $L_{\textit{content}} = \textit{CTCLoss}(p_c, y_c)$
\noindent \textbf{2) Speaker Verification (SV):} The speaker verification head also comprises two linear layers. Since the entire utterance should map to a single speaker representation, we map the Conformer output at the first time-step to a speaker embedding $z_s = f_{\theta_{s1}} ( {z}_1 ) $. 
Finally, a second linear layer maps the speaker embedding to scores over all the speakers in the dataset, that is $p_s = f_{\theta_{s2}} ( z_s ) $. The SV head is trained on audio-speaker pairs $(x, y_s)$ using angular softmax loss~\cite{Liu_2017_CVPR}, that is, $L_{SV} = \textit{AngularSoftMax}(p_s, y_s)$ \\\\
\noindent\textbf{Multi-task Training: }
Since large speaker-verification datasets like VoxCeleb~\cite{voxcelebwildjournal} do not have text transcripts, we use separate datasets and data loaders for each task. 
During each training iteration, we load the mini-batches for both tasks and compute the task-specific losses. Next, we combine the two losses to obtain $L_{\textit{multi}} = L_{\textit{content}} + \alpha L_{\textit{SV}}$ and backpropagate through both the downstream heads and the Conformer backbone to update the parameters. 
We use Adam optimizer with a learning rate of $10^{-4}$ for the downstream network parameters and $10^{-5}$ for the Conformer backbone. We use a lower learning rate for Conformer to prevent over-fitting during finetuning.\\

\begin{figure}[htp]
    \centering
    \includegraphics[width=1.0\columnwidth]{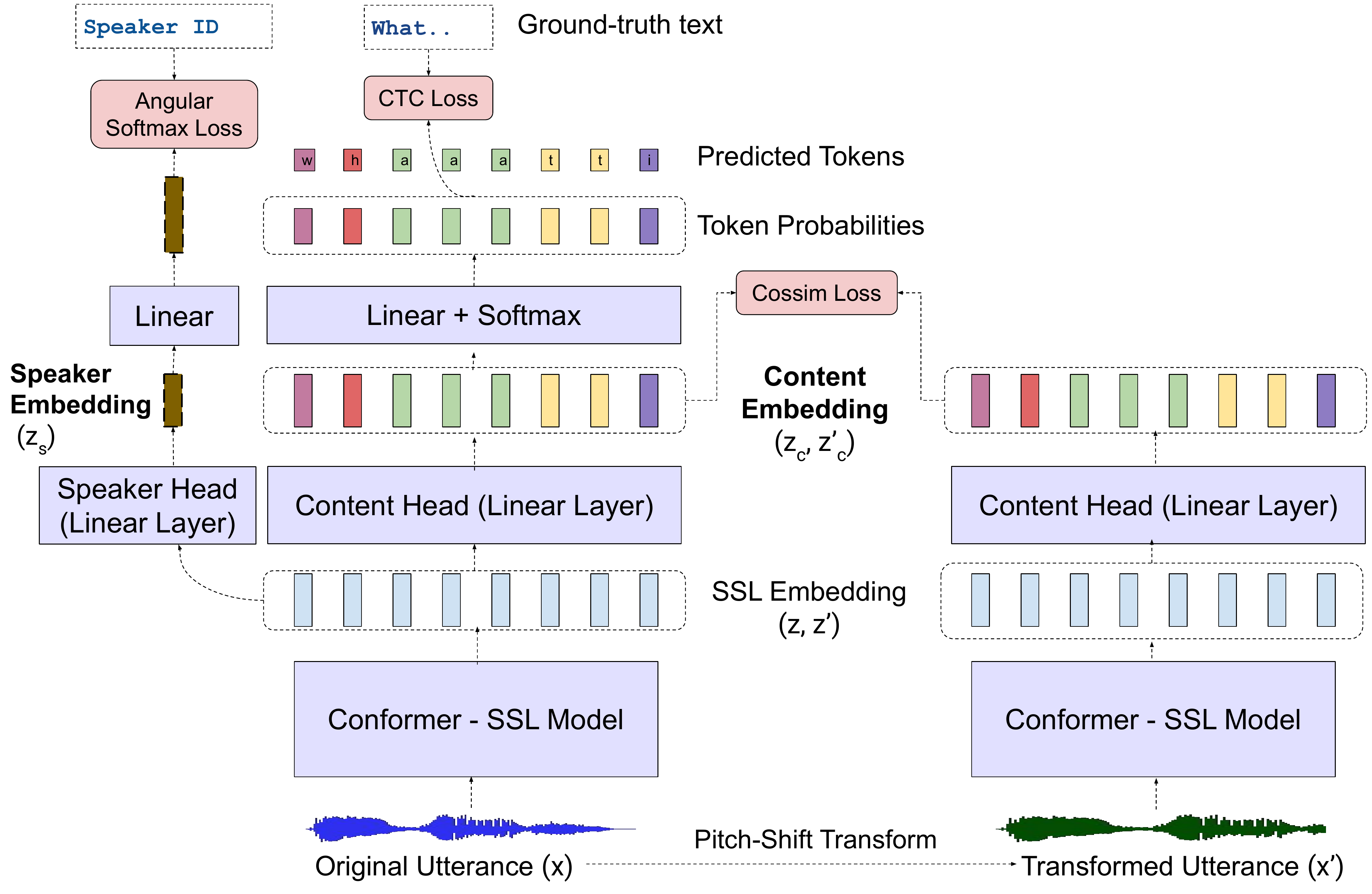}
    \vspace{-8mm}
    \caption{Speech Representation Extractor (SRE) training
    }
    \vspace{-6mm}
    \label{figs:conformer_diagram}
\end{figure}

\noindent\textbf{Disentangling content and speaker representations: }
In our initial experiments (Section~\ref{sec:disentangle}), we find that speaker information can leak into the linguistic content embedding $z_c$ if we simply perform the multi-task training as described above.
This information leak is not surprising since the underlying SSL encodings $z$ are being fine-tuned to contain both speaker and linguistic content information. 
Without proper disentanglement, the synthesis network largely ignores the speaker embedding $z_s$ thereby not allowing for effective voice conversion.
To address this challenge, we synthetically alter the voice of a given utterance $x$ using pitch-shift data augmentation during training and obtain a pitch-shifted audio $x'$.
Next, we use a Siamese cosine similarity loss to encourage similarity between the content embeddings of the original ($z_c=f_{\theta_{c1}} (E(x))$) and pitch-shifted audio ($z_c' = f_{\theta_{c1}} (E(x')) $ as follows:

\begin{equation}
   L_{\textit{disentangle}} = 1 - S_{\textit{cos}} (z_c, z_c')  
\label{eq:disentangle}
\end{equation}

where $S_{\textit{cos}}$ is the cosine similarity between two vectors. 
The above disentanglement loss is added to our multi-task training loss to obtain $L_{\textit{SRE}} = L_{\textit{multi}} + \beta L_{\textit{disentangle}} $ 
The above loss term encourages content embeddings to be independent of the voice or the speaker thereby achieving our goal of disentanglement. Figure~\ref{figs:conformer_diagram} gives an overview of SRE training.

\subsection{Mel-Spectrogram Synthesizer}
The task of the synthesizer is to reconstruct the ground-truth mel-spectrogram from the representations given by the SRE. 
Note that the temporal characteristics and length of the content representation $z_c$ determines the speaking rate and total duration of the given utterance. Therefore, a synthesizer trained on the raw output of the SRE can simply determine the speaking rate from $z_c$ and not adapt the speaking rate for a different target speaker-embedding $z_s'$. This limitation exists in many past works~\cite{polyak2021speech,choi2021neural} in the voice conversion domain where the duration and speaking rate of the voice-swapped audio are always the same as that of the source audio. 

To address the above challenge, we first process the raw content representation $z_c$, by grouping together consecutive vectors that have the same predicted token. Consider consecutive content vectors ${z_c}_i \dots {z_c}_j$ from time-steps $i$ through $j$ that have the same predicted language token, that is $\textit{argmax}({p_c}_i) = \dots = \textit{argmax}({p_c}_j)$. We group these consecutive vectors and average them along the temporal dimension to obtain the new vector ${g_c}_t = \textit{average}({z_c}_i \dots {z_c}_j)$ at a time-step $t$. The target ground-truth duration for ${g_c}_t$ is determined by the number of grouped vectors, that is, ${d_c}_t = j - i + 1$. Therefore, by repeating this procedure for all time-steps, we obtain the grouped content representation and target durations as $g_c, d_c = G(z_c)$.

\begin{figure}[htp]
    \centering
    \includegraphics[width=1.0\columnwidth]{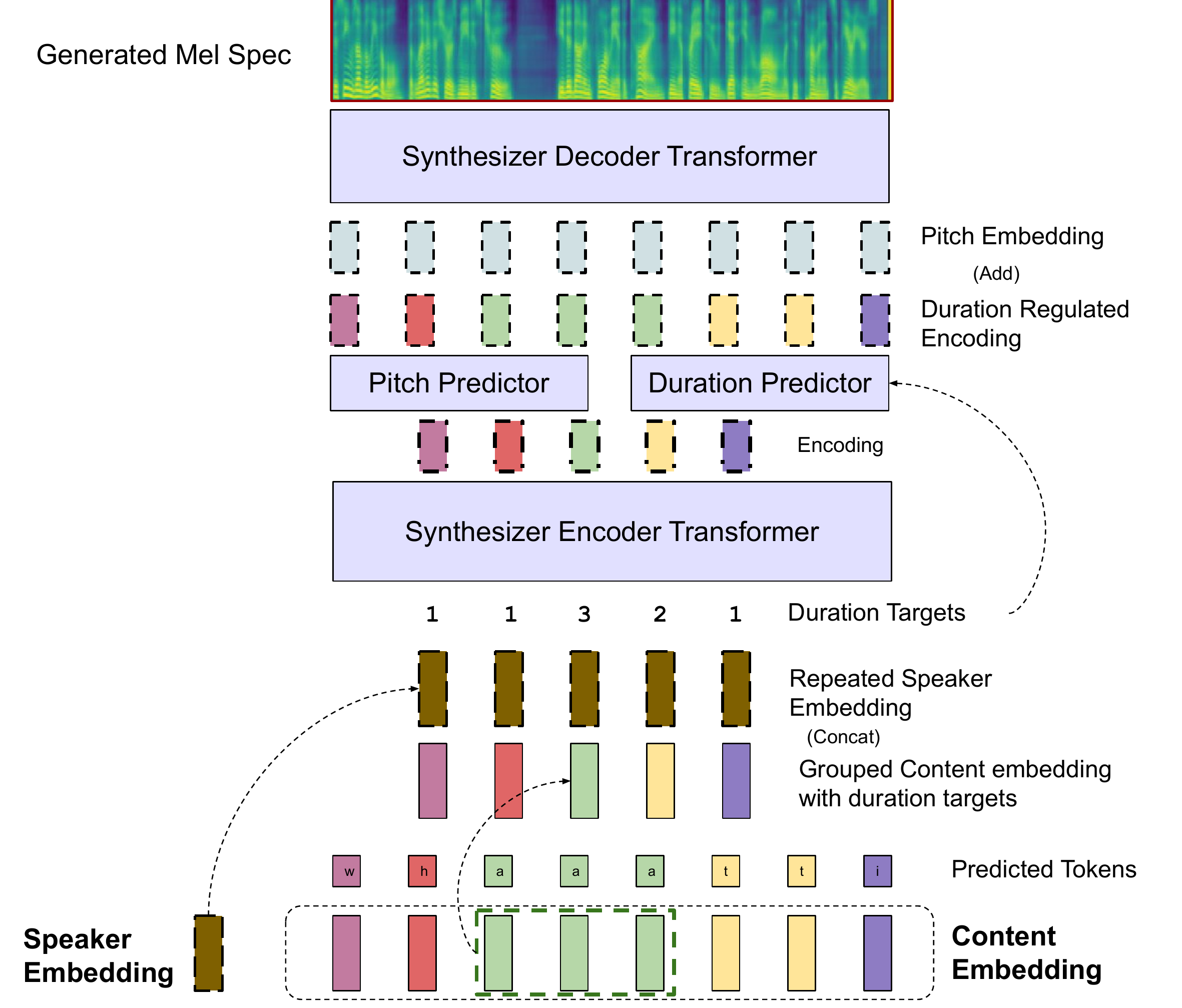}
    \vspace{-8mm}
    \caption{Mel-spectrogram synthesis from SRE representations}
    \label{figs:synthesizer_diagram}
\end{figure}

Next, we model the duration $d_c$ and fundamental frequency $p$ (pitch contour) of the audio as a function of the processed content embedding $g_c$ and speaker embedding $z_s$. To achieve this, our synthesizer network consists of two feed-forward transformers $F_e$ and $F_d$ similar to FastPitch~\cite{lancucki2021fastpitch}, but operates on input $g_c$ and $z_s$ instead of text. The hidden representation from the first transformer is used to predict the duration and pitch. That is, $h=F_e(g_c, z_s)$. $\hat{d} = \textit{DurationPredictor}(h), \hat{p} = \textit{PitchPredictor}(h)$. Next, the pitch contour is projected and averaged over each time-step of the hidden representation $h$ and added to $h$ to get $k = h + \textit{PitchEmbedding}(p)$. Finally, $k$ is discretely upsampled as per the ground-truth duration $d_c$ and fed as input to the second transformer to get the predicted mel-spectrogram $\hat{y} = F_d( \textit{DurationRegulation}(k, d_c) )$. Figure~\ref{figs:synthesizer_diagram} describes the synthesizer model. 

The synthesizer is trained on a multi-speaker dataset in a text-free manner. The ground-truth pitch contour $p$ is derived from the Yin algorithm~\cite{de2002yin}. The pitch contour of each utterance is normalized using the mean and standard deviation of the pitch contours of the given speaker. This per-speaker normalization ensures that the pitch contour only captures the prosodic changes over time and not the speaker's identity. The model is trained to optimize three losses --- mel-reconstruction error, pitch prediction error and duration prediction error. That is,
\begin{equation}
  L_\textit{synth} = \lVert \hat{{y}} - {y}\rVert^2_2 + 
    \lambda_1      \lVert \hat{{p}} - {p}\rVert^2_2 + 
    \lambda_2      \lVert \hat{{d}} - {d_c}\rVert^2_2 
\end{equation}






\section{Experiments and Results}

\subsection{Datasets and Training}
\textbf{SRE:} We initialize the backbone of our SRE with the weights of a Conformer model trained using self-supervised learning on the entire LibriSpeech dataset~\cite{librispeechref} containing 960 hours of English speech. 
To train the SRE in multi-task setting, we use the \textit{train-clean-360} subset of LibriSpeech~\cite{librispeechref} for speech recognition (with characters as the language tokens) and  VoxCeleb-2~\cite{voxcelebwildjournal} dataset for speaker verification. 
For speech recognition, we use utterances with lengths between $4$ seconds to $16$ seconds. For speaker verification, we use two-second utterance slices in both training and inference. All utterances are resampled to $22050$ Hz and mel-spectrograms are extracted with $80$ bands using FFT size=$1024$, window size=$1024$, and hop size=$256$. With these hyper-parameters, $46$ ms of speech map to one content vector. 
We evaluate the speech recognition performance on the LibriSpeech test set and speaker-verification performance on the original trial pairs of VoxCeleb-1~\footnote{https://www.robots.ox.ac.uk/~vgg/data/voxceleb/vox1.html}. Our multi-task SRE achieves a Character Error Rate (CER) of 2.9\% for speech recognition and an Equal Error Rate (EER) of 3.01\% for speaker verification.

\noindent \textbf{Synthesizer:} The mel-spectrogram synthesizer and the HiFiGAN vocoder are trained on the \textit{train-clean-360} subset of LibriTTS 
using the same STFT parameters as the SRE. We point the readers to our codebase linked in the first page for implementation and hyperparameter details.

\subsection{Feature Disentanglement}
\label{sec:disentangle}
To evaluate how effectively we can remove speaker information from the content representation, we train a speaker classifier on just the content representation. Lower speaker classification accuracy indicates more effective disentanglement. 
We perform an ablation study by training three speaker classifiers using three different representations as input: 1) Speaker representations, 2) Content representations trained without $L_{\textit{disentangle}}$, 3) Content representations trained with  $L_{\textit{disentangle}}$. Each speaker classifier is a three-layer neural network with $256$ hidden units per hidden layer. The classifiers are trained on $40$ speakers from the dev-clean subset of LibriTTS and evaluated on $5$ unseen utterances of each speaker.  
As shown by the results reported in Table~\ref{tab:disentanle}, a classifier trained with content representation using the $L_{\textit{disentangle}}$ 
given by Equation~\ref{eq:disentangle}, 
achieves the lowest speaker classification accuracy while still achieving low CER on text recognition. This result indicates the effective removal of speaker information from content representation without compromising its ability to capture linguistic features.

\setlength{\tabcolsep}{4pt}
\begin{table}[t]
\centering

\resizebox{0.9\columnwidth}{!}{%
\begin{tabular}{l|ccc}
\multicolumn{1}{c}{} & \multicolumn{3}{c}{\emph{Embedding Type}} \\
\toprule
& Speaker Emb. & Content Emb. & Content Emb. \\
&  &  (w/o $L_\textit{disentangle})$ & (with $L_\textit{disentangle}$)  \\
\midrule
Spk-class. Acc. & 97.5\% & 62.5\% & \textbf{37.5\%} \\
ASR CER  & - & 2.9\% & 2.9\%  \\
\bottomrule
\end{tabular}
}
\vspace{-2mm}
\caption{\footnotesize{Evaluating feature disentanglement. 
Lower speaker-classification accuracy for content embeddings trained with disentanglement loss indicates effective removal of speaker information from the content embedding.}}
\vspace{-4mm}
\label{tab:disentanle}
\end{table}

\subsection{Voice conversion}
To perform voice conversion, we use the synthesis model to combine the content embedding of any given source utterance with the speaker embedding of the target speaker, both of which are derived from the SRE. The target speaker embedding is estimated from $10$ seconds of speech of the given speaker.
We consider two voice conversion scenarios - for a seen speaker to another seen speaker (Many-to-Many) and from an unseen speaker to another unseen speaker (Any-to-Any). For seen speakers, we use the holdout utterances of the \textit{train-clean-360} subset of LibriTTS dataset, and for unseen speakers, we use  the \textit{dev-clean} subset. For each scenario, we randomly select $20$ target speakers ($10$ male and $10$ female). Next, we select $10$ source utterances, each one from $10$ alternate speakers. This results in a total of $200$ voice conversion trials in each scenario. 

We evaluate the synthesized speech on three aspects -- speaker-similarity, intelligibility, and naturalness. For speaker similarity, we compute the speaker embeddings of synthesized and real utterances using a separate SV model~\cite{koluguri2020speakernet}. 
Then we pair the synthesized and real utterances to create an equal number of positive and negative pairs for each target speaker (a total of $4000$ pairs for each technique) to compute the Equal Error Rate (SV-EER). 
Automatic SV metrics are popularly used by prior work~\cite{huang2022s3prl,gu2021mediumvc,lin2021fragmentvc} to evaluate voice conversion methods and have shown a strong correlation with human perception of speaker similarity.
For intelligibility, we transcribe the generated and the source utterances using a pre-trained ASR model~\cite{kriman2020quartznet} and compute the mean character error rate between the two. For naturalness, we conduct a mean-opinion-score (MOS) study on Amazon Mechanical Turk where we ask the listeners to rate each utterance on a scale of 1 to 5. Each utterance is rated by $2$ independent listeners resulting in $400$ evaluations of each technique.

We compare our framework against several prior voice conversion methods on the same training and test benchmark as ours using their official open-source implementations.
As shown in the results reported in Table~\ref{tab:vc}, our model outperforms prior work on all three metrics. As evident by the lower SV-EER in Table~\ref{tab:vc}, our best-performing model generates voice-converted speech with high speaker-similarity to our target speaker with improved naturalness and intelligibility. We encourage readers to listen to our audio examples linked in the first page.

\setlength{\tabcolsep}{2pt}
\begin{table}[t]
\centering
\resizebox{\columnwidth}{!}{%
\begin{tabular}{l|rrr|rrr}
\multicolumn{1}{c}{} & \multicolumn{3}{c}{\emph{Many-to-Many}} & \multicolumn{3}{c}{\emph{Any-to-Any}} \\
\toprule
Technique & SV-EER. & CER & {MOS } &  SV-EER. & CER & {MOS } \\
\midrule
Real Data & $4.3\%$ & - & $4.01 \pm 0.09$& $4.2\%$ & - & $4.04 \pm 0.09$ \\
\midrule
AutoVC~\cite{qian2019autovc} & $24.3\%$ & $20.4\%$ & $2.87 \pm 0.12$& $36.3\%$ & $35.4\%$ & $2.56 \pm 0.12$\\
AdaIN-VC~\cite{chou2019one} & $17.4\%$ & $27.4\%$ & $2.60 \pm 0.12 $ & $25.8\%$ & $29.7\%$ & $2.76 \pm 0.11$\\
MediumVC~\cite{gu2021mediumvc}  & $10.5\%$ & $32.1\%$ & $3.12 \pm 0.13$ & $21.4\%$ & $35.4\%$ & $3.09 \pm 0.12$ \\
FragmentVC~\cite{lin2021fragmentvc} & $16.6\%$ & $28.1\%$ & $3.22 \pm 0.11$ & $23.5\%$ & $39.7\%$ & $3.05 \pm 0.12$\\
S3PRL-VC~\cite{huang2022s3prl} & $13.5\%$ & $9.2\%$ & $3.07 \pm 0.11$ & $21.3\%$ & $8.9\%$ & $3.17 \pm 0.13$ \\
YourTTS~\cite{casanova2022yourtts}  & $9.5\%$ & $5.8\%$ & $3.59 \pm 0.09$& $13.4\%$ & $5.3\%$ & $3.60 \pm 0.09$\\
\midrule
ACE-VC (Ours) & $\mathbf{5.5\%}$ & $\mathbf{2.7\%}$ & $\mathbf{3.62 \pm 0.10}$& $\mathbf{8.4\%}$ & $\mathbf{2.8\%}$ & $\mathbf{3.75 \pm 0.09}$\\
\bottomrule
\end{tabular} 
}
\vspace{-3mm}
\caption{\footnotesize{Comparison of different voice-conversion techniques. Lower values for SV-EER and CER are desirable for higher speaker similarity and intelligibility respectively. Higher MOS (reported with 95\% confidence interval) indicates more natural-sounding speech.}}
\vspace{-4mm}
\label{tab:vc}
\end{table}

\vspace{-2mm}
\section{Conclusion}
In this work, we propose a zero-shot voice conversion framework using disentangled speech representations. We propose a speech representation extractor that effectively and explicitly disentangles content and speaker information using a single Conformer-SSL backbone, without quantizing or compressing the learned representations. Our synthesis model can predict the speaking rate and pitch contour from both the speaker and content embeddings resulting in a controllable and adaptive voice conversion model. Our results indicate a significant improvement over recently proposed and state-of-the-art voice conversion models in terms of speaker similarity, naturalness, and intelligibility of generated speech. 

\bibliographystyle{IEEEbib}
\bibliography{bib}

\begin{thebibliography}{10}

\bibitem{saito2011one}
D.~Saito, K.~Yamamoto, N.~Minematsu, and K.~Hirose,
\newblock ``One-to-many voice conversion based on tensor representation of
  speaker space,''
\newblock in {\em Interspeech}, 2011.

\bibitem{mohammadi2017overview}
S.~H. Mohammadi and A.~Kain,
\newblock ``An overview of voice conversion systems,''
\newblock in {\em Speech Communication}. Elsevier, 2017.

\bibitem{chou2019one}
J.~Chou and H.~Y. Lee,
\newblock ``One-shot voice conversion by separating speaker and content
  representations with instance normalization,''
\newblock {\em Interspeech}, 2019.

\bibitem{qian2019autovc}
K.~Qian, Y.~Zhang, S.~Chang, X.~Yang, and M.~Hasegawa-Johnson,
\newblock ``Autovc: Zero-shot voice style transfer with only autoencoder
  loss,''
\newblock in {\em ICML}. PMLR, 2019.

\bibitem{choi2021neural}
H.~S. Choi, J.~Lee, W.~Kim, J.~Lee, H.~Heo, and K.~Lee,
\newblock ``Neural analysis and synthesis: Reconstructing speech from
  self-supervised representations,''
\newblock {\em NeurIPS}, 2021.

\bibitem{casanova2022yourtts}
E.~Casanova, J.~Weber, C.D. Shulby, A.C. Junior, E.~G{\"o}lge, and M.~A. Ponti,
\newblock ``Yourtts: Towards zero-shot multi-speaker tts and zero-shot voice
  conversion for everyone,''
\newblock in {\em ICML}. PMLR, 2022.

\bibitem{sun2016phonetic}
L.~Sun, K.~Li, H.~Wang, S.~Kang, and H.~Meng,
\newblock ``Phonetic posteriorgrams for many-to-one voice conversion without
  parallel data training,''
\newblock in {\em ICME}, 2016.

\bibitem{tian2018average}
X.~Tian, J.~Wang, H.~Xu, E.~S. Chng, and H.~Li,
\newblock ``Average modeling approach to voice conversion with non-parallel
  data.,''
\newblock in {\em Odyssey}, 2018.

\bibitem{lakhotia2021generative}
K.~Lakhotia, E.~Kharitonov, W.N. Hsu, Y.~Adi, A.~Polyak, B.~Bolte, T.~Nguyen,
  J.~Copet, A.~Baevski, A.~Mohamed, et~al.,
\newblock ``On generative spoken language modeling from raw audio,''
\newblock {\em Transactions of the Association for Computational Linguistics},
  2021.

\bibitem{polyak2021speech}
A.~Polyak, Y.~Adi, J.~Copet, E.~Kharitonov, K.~Lakhotia, W.N. Hsu, A.~Mohamed,
  and E.~Dupoux,
\newblock ``Speech resynthesis from discrete disentangled self-supervised
  representations,''
\newblock in {\em Interspeech}, 2021.

\bibitem{lin2021fragmentvc}
Y.~Lin, C.~M. Chien, J.~H. Lin, H.~Lee, and L.~S. Lee,
\newblock ``Fragmentvc: Any-to-any voice conversion by end-to-end extracting
  and fusing fine-grained voice fragments with attention,''
\newblock in {\em ICASSP}. IEEE, 2021.

\bibitem{huang2022s3prl}
W.~C. Huang, S.~W. Yang, T.~Hayashi, H.~Y. Lee, S.~Watanabe, and T.~Toda,
\newblock ``S3prl-vc: Open-source voice conversion framework with
  self-supervised speech representations,''
\newblock in {\em ICASSP}. IEEE, 2022.

\bibitem{wav2vec2}
A.~Baevski, Y.~Zhou, A.~Mohamed, and M.~Auli,
\newblock ``wav2vec 2.0: A framework for self-supervised learning of speech
  representations,''
\newblock {\em NeurIPS}, 2020.

\bibitem{gulati2020conformer}
A.~Gulati, J.~Qin, C.~Chiu, N.~Parmar, Y~Zhang, J.~Yu, W.~Han, S.~Wang,
  Z.~Zhang, Y.~Wu, et~al.,
\newblock ``Conformer: Convolution-augmented transformer for speech
  recognition,''
\newblock {\em Interspeech}, 2020.

\bibitem{hussain2022multi}
S.~Hussain, V.~Nguyen, S.~Zhang, and E.~Visser,
\newblock ``Multi-task voice activated framework using self-supervised
  learning,''
\newblock in {\em ICASSP}. IEEE, 2022.

\bibitem{bromley1993signature}
J.~Bromley, I.~Guyon, Y.~LeCun, E.~S{\"a}ckinger, and R.~Shah,
\newblock ``Signature verification using a" siamese" time delay neural
  network,''
\newblock {\em NIPS}, 1993.

\bibitem{kong2020hifi}
J.~Kong, J.~Kim, and J.~Bae,
\newblock ``Hifi-gan: Generative adversarial networks for efficient and high
  fidelity speech synthesis,''
\newblock {\em NeurIPS}, 2020.

\bibitem{graves2006connectionist}
A.~Graves, S.~Fern{\'a}ndez, F.~Gomez, and J.~Schmidhuber,
\newblock ``Connectionist temporal classification: labelling unsegmented
  sequence data with recurrent neural networks,''
\newblock in {\em ICML}. ACM, 2006.

\bibitem{Liu_2017_CVPR}
W.~Liu, Y.~Wen, Z.~Yu, M.~Li, B.~Raj, and L.~Song,
\newblock ``Sphereface: Deep hypersphere embedding for face recognition,''
\newblock in {\em CVPR}. IEEE, 2017.

\bibitem{voxcelebwildjournal}
A.~Nagrani, J.S. Chung, W.~Xie, and A.~Zisserman,
\newblock ``Voxceleb: Large-scale speaker verification in the wild,''
\newblock {\em Computer Speech \& Language}, 2020.

\bibitem{lancucki2021fastpitch}
A.~{\L}a{\'n}cucki,
\newblock ``Fastpitch: Parallel text-to-speech with pitch prediction,''
\newblock in {\em ICASSP}. IEEE, 2021.

\bibitem{de2002yin}
A.~De~Cheveign{\'e} and H.~Kawahara,
\newblock ``Yin, a fundamental frequency estimator for speech and music,''
\newblock in {\em The Journal of the Acoustical Society of America}, 2002.

\bibitem{librispeechref}
V.~Panayotov, G.~Chen, D.~Povey, and S.~Khudanpur,
\newblock ``Librispeech: an asr corpus based on public domain audio books,''
\newblock in {\em ICASSP}. IEEE, 2015.

\bibitem{koluguri2020speakernet}
N.R. Koluguri, J.~Li, V.~Lavrukhin, and B.~Ginsburg,
\newblock ``Speakernet: 1d depth-wise separable convolutional network for
  text-independent speaker recognition and verification,''
\newblock {\em arXiv:2010.12653}, 2020.

\bibitem{gu2021mediumvc}
Y.~Gu, Z.~Zhang, X.~Yi, and X.~Zhao,
\newblock ``Mediumvc: Any-to-any voice conversion using synthetic
  specific-speaker speeches as intermedium features,''
\newblock {\em arXiv:2110.02500}, 2021.

\bibitem{kriman2020quartznet}
S.~Kriman, S.~Beliaev, B.~Ginsburg, J.~Huang, O.~Kuchaiev, V.~Lavrukhin,
  R.~Leary, J.~Li, and Y.~Zhang,
\newblock ``Quartznet: Deep automatic speech recognition with 1d time-channel
  separable convolutions,''
\newblock in {\em ICASSP}. IEEE, 2020.

\end{thebibliography}

\end{document}